\documentclass{article}

\usepackage{amsmath}
\usepackage{graphicx}
\usepackage{natbib}
\usepackage{amsfonts}

\newsavebox{\astrutbox}
\sbox{\astrutbox}{\rule[-5pt]{0pt}{20pt}}

\begin{document}

\title{On the inconsistency of the Camassa-Holm model with the shallow water theory}
\author{Rikesh Bhatt and Alexander V Mikhailov}
\date{}
\maketitle

\begin{abstract}
In our paper we show that the Camassa-Holm equation does not represent a long wave asymptotic due to a major  inconsistency with the theory of shallow water waves.  We state that any solution of the Camassa-Holm equation, which is not asymptotically close to a solution of the Korteweg--de Vries equation is an artefact of the 
model and irrelevant to the theory of shallow water waves.

\end{abstract}

\section{Introduction}
A partial differential equation
\begin{equation}\label{cceq0}
2\,\omega\,U_{{y}}+U_{{\tau}}+3\,U_{{y}}U-U_{{\tau yy}}-2\,U_{y}U_{yy}-UU_{yyy}=0 
\end{equation}
known in the literature as Camassa-Holm equation is a fine example of an integrable  system  with many interesting and rather 
unusual properties. As an integrable equation it was discovered in \cite{ff0}, \cite{ff1}, but equation (\ref{cceq0}) had not been really noticed until the publication  \cite{CH} where the authors proposed it as a model for shallow water waves. Moreover, Camassa and Holm have 
shown that in the case $\omega=0$ equation (\ref{cceq0}) possesses a peculiar solution with a cusp 
\begin{equation}\label{peakon}
 U=v\exp(- |y-vt|)
\end{equation}
which they called a ``peakon''. Moreover, they also found exact multi-peakon solutions. The paper of Camassa and Holm has triggered 
an enormous avalanche of publications with the developments of mathematical theory for this new class of integrable systems and  speculations about possible applications of the Camassa-Holm equation to the theory of shallow water waves (including the problem of tsunami, \cite{laksh}) and other long wave asymptotic theories. 

There were a number of publications where  authors 
raised some criticism to the original derivation and proposed new versions of  the derivations starting from the reduction of the Green-Naghdi model, reduction of the generalised Serre model or a direct multiscaling
asymptotic expansion of the Euler equation (see for example \cite{johnson2000}, \cite{dullinetal}, \cite{constantinlannes09}, \cite{dias} \cite{camassaetal94}). The main concern of these  derivations was to achieve a fine adjustment of the coefficients in the first few terms of the asymptotic expansion in order to make  the equation corresponding to the truncated expansion integrable. The aim of this paper is to point out that the Camassa-Holm equation does not correspond to any dominant balance in the asymptotic expansion and thus it does not represent a long wave asymptotic for water waves. We have analysed the above derivations and found them inconsistent with the basic principles of asymptotic theory. Indeed, one cannot keep the first principal asymptotic contribution (corresponding to the Korteweg--de Vries theory) together with the next asymptotic correction, then truncate the expansion and balance these two contributions by a re-scaling. Such a re-scaling contains the small asymptotic parameter and thus violates the original assumption about the ratio of the water depth and the characteristic wave length. We have computed the next (neglected in the above derivations) term of the expansion and have shown that after the proposed re-scaling it is of the same order as any term accounted in the Camassa-Holm equation. Moreover, the parameter $\omega$ in (\ref{cceq0}) appears in denominators of the neglected terms and thus cannot be set to zero (for the existence of peakon solutions) in any long-wave asymptotic theory. Here we should mention that the fact that the peakon solutions are irrelevant for shallow water waves was well understood earlier and published in \cite{johnson2000}, \cite{dullinetal}. Using the exact soliton solution  of the Camassa-Holm equation we have shown that the neglected terms are of the same order as  terms accounted in equation (\ref{cceq0}).  

\section{Long wave expansion and the Camassa-Holm Equation}

In this section, following \cite{whitham74}, we give a sketch of  long wave asymptotic expansion beyond the Korteweg--de Vries (KdV) theory . We shall illustrate the derivation of the Camassa-Holm equation following \cite{dullinetal}, \cite{holmdull} and point out where the inconsistency occurs. Also we will show that the terms neglected in the theory of the Camassa-Holm equation are of the same order as any term accounted in (\ref{cceq0}). We claim that the inconsistency cannot be removed if one uses a reduction of the Green-Naghdi model (which itself is an approximation), or by choosing the value  of the velocity potential inside of the flow (\cite{johnson2000}) or by any other method. We shall neglect the surface tension, since it does not affect our argument but simply makes expressions look more complicated. Also we shall neglect viscosity and compressibility of water.

Two dimensional motion of vorticity free fluid is described by the potential $\Phi(x',z',t')$ of the velocity field ${\bf u}(x',z',t')={\bf \nabla} \Phi$.  In the bulk of the fluid the potential satisfies the Laplace equation
\[
 \Phi_{x'x'}+\Phi_{z'z'}=0
\]
with the boundary condition $\Phi_{z'}=0$ at the bottom $z'=0$. At the free surface of the fluid $z'=h_0+H(x',t')$   there are kinematic and dynamic boundary conditions 
\begin{eqnarray*}
&&\Phi_{z'}=  H_{x'}\Phi_{x'} +  H_{t'}  \\
&&\Phi_{t'} + \frac{1}{2}\left(  \Phi^{2}_{x'} + \Phi^{2}_{z'} \right) + gH=0  ,
\end{eqnarray*}
where $g$ is the acceleration due to gravity and $h_0$ is the undisturbed depth of water.

Long wave asymptotic expansion assumes a small parameter $\epsilon=h_0^2/L^2$ where $L$ is a typical wavelength of the wave. Another dimensionless small parameter of the theory is $\mu=a_0/h_0$ where $a_0$ is a typical amplitude of the wave.  We shall assume that $\epsilon\le \mu\ll 1$.
Actually one can set $\mu=\epsilon$ and develop the theory with one parameter, but following \cite{johnson2000}, \cite{dullinetal}, \cite{constantinlannes09} we shall keep both parameters for better control over the terms.

Introducing  dimensionless variables
\[
x' = Lx ,\  z' = h_{0}z,\  t' = \frac{L}{\sqrt{gh_{0}}} t,\  
H= \mu h_0 \eta ,\ \Phi = \mu l \sqrt{gh_{0}} \phi ,\ 
\]
we re-write the above system of equations in the form:
\begin{eqnarray}
&&\epsilon\phi_{xx} + \phi_{zz}= 0\label{Laplace}\\&& \left[\phi_{z}\right]_{z=0}=0 \label{bottom} \\ 
&&\left[\frac{1}{\epsilon}\phi_{z}- \mu \eta_{x}\phi_{x} -  \eta_{t}\right]_{z=1 + \mu\eta(x,t)}=0 \label{kinematic} \\
&&\left[\phi_{t} + \frac{1}{2}\left( \mu\phi^{2}_{x} + \frac{\mu}{\epsilon}\phi^{2}_{z} \right) +  \eta\right]_{z=1 + \mu\eta(x,t)} = 0  \label{dynamic} 
\end{eqnarray}

Starting from here we shall develop asymptotic expansion as a series in $\epsilon^n\mu^m,\ n,m\in{\mathbb{Z}}_+$. We shall illustrate the derivation the Camassa-Holm equation with corrections in three steps. We begin with the derivation of  the Boussinesq expansion up to the order $\epsilon^n\mu^m,\ n+m= 3$. Then, following  \cite{whitham74}, we make a reduction to the KdV theory describing unidirectional wave propagation.  Finally we transform the equation obtained to the form (\ref{cceq0}), keeping terms of order $\epsilon^n\mu^m,\ n+m= 3$ to  demonstrate  that the Camassa-Holm equation does not represent a long wave asymptotic for surface waves.   

\subsection{The Boussinesq expansion}

The Boussinesq expansion aims to eliminate the dependence on the vertical coordinate $z$ and reduce  (\ref{Laplace}) -
(\ref{dynamic}) to a system of equations on the elevation $\eta=\eta(x,t)$ and the horizontal component of the velocity field at the bottom $w=\phi(x,0,t)_x$. In this Section we shall follow the construction presented in detail in \cite{whitham74} (Chapter 13.11), but will keep more terms in the expansion.

It follows from the Laplace equation (\ref{Laplace}) and the boundary condition at the bottom (\ref{bottom}) that 
\begin{equation}\label{phiexp}
\phi(x,z,t)=\sum_{n=0}^{\infty}\epsilon^n(-1)^n\frac{z^{2n}}{(2n)!}\frac{\partial^{2n}F}{\partial x^{2n}} ,
\end{equation}
where $F(x,t)$ is the value of the potential $\phi$ at the bottom\footnote{In \cite{dullinetal} the authors use a different geometry, namely the bottom is set at $z=-1$. Their solution  does not satisfy the boundary condition at $z=-1$ (presumably due to a misprint). To rectify the misprint one has to  replace $z$  by $z+1$ in the right hand side of (2.7),(2.8) in \cite{dullinetal}.}, and thus $w=F_x$.

In order to reduce the system (\ref{Laplace}) - (\ref{dynamic}) to two equations for functions $\eta(x,t)$ and $w(x,t)$ we substitute $\phi(x,z,t)$ (\ref{phiexp}), in (\ref{kinematic}) and (\ref{dynamic}), then we differentiate in $x$ the equations obtained from (\ref{dynamic}) and replace $F_x$ by $w$ in the both equations. Keeping terms of order $\epsilon^n\mu^m,\ n+m\le 3$ we get
\begin{eqnarray}
&&0 = \eta_{{t}}+w_{{x}}+ \mu\left( \eta w  \right)_{{x}} -\frac{\epsilon}{6}w_{{xxx}} \nonumber \\
&&-\frac{ \mu\,\epsilon}{2}\left(\eta w_{{xx}}\right)_{{x}}+{\frac {{\epsilon}^{2}}{120}} w_{{xxxxx}}  \label{bc1} \\ 
&&-\frac{{\mu}^{2}\epsilon}{2}\left(w_{{xx}}{\eta}^{2} \right)_x  +\nonumber \frac{\mu\,{\epsilon}^{2}}{24}\left(w_{{xxxx}} \eta \right)_x -{\frac {{\epsilon}^{3}}{5040}}\,w_{{xxxxxxx}} 
\end{eqnarray}
and
\begin{eqnarray}
&&0 = w_{{t}}+\eta_{{x}}+\mu\,w w_{{x}}-\frac{\epsilon}{2} w_{{txx}} \nonumber \\
&& -\frac{\mu\,\epsilon}{2}\left( 2w_{{tx}}\eta+w w_{{xx}}-w_{{x}}^2\right)_x + \frac{{\epsilon}^{2}}{24}w_{{txxxx}}  \label{bc2}  \\ 
&&-\frac{ {\mu}^{2}\epsilon}{2} \left(2w w_{{xx}}\eta -2{w_{{x}}}^{2}\eta+w_{{tx}}{\eta}^{2} \right)_x \nonumber \\ 
&&+ \frac{\mu\,{\epsilon}^{2}}{24}\left( 4\,w_{{txxx}}\eta+3\,w_{{xx}}^2-4\,w_{{x}}w_{{xxx}}+w_{{xxxx}}w  \right)_x 
-{\frac {{\epsilon}^{3}}{720}}\,w_{{txxxxxx}} \nonumber
\end{eqnarray}
respectively. First two lines in (\ref{bc1}),(\ref{bc2}) coincide with equation (2.9) in \cite{dullinetal}\footnote{In \cite{dullinetal} in the first equation (2.9) there is a misprint in the sign at the term proportional to $\delta^4$ (in our paper it is the term proportional to $\epsilon^2$ in (\ref{bc1})).}. For the purpose of our paper we are keeping the next order in the expansion. There are no obstructions to find higher order terms if required.

\subsection{Reduction To Unidirectional Waves. The KdV theory with higher asymptotic corrections} \label{uniwaves}

System (\ref{bc1}),(\ref{bc2}) describes waves propagating in both direction. There are many ways to reduce it to unidirectional wave propagation. In order to be consistent with \cite{dullinetal} we employ the method proposed in \cite{whitham74}. Namely, we shall assume  
$$w=\eta+\sum_{k=1}^\infty \sum_{n=0}^k \mu^n\epsilon^{k-n}f_{kn}[\eta]$$ 
and request that equations (\ref{bc1}),(\ref{bc2}) coincide upon this assumption, that would enable us to determine the coefficients $f_{kn}[\eta]$. Keeping terms with $k\le 3$ we get

\begin{multline}
w = \indent \eta  -\,\frac{\mu}{4}\, \eta^2 +\,\frac{\epsilon}{3}\, \eta_{xx}\\
 +\frac{{\mu}^{2}}{8}\, \eta^3+\frac{\epsilon\,\mu}{16}\left(3 \,\eta_x^2 +  8\, \eta\eta_{xx}\right) +\frac{{\epsilon}^{2}}{10}\, \eta_{xxxx}\\ - \frac{5\,{\mu}^{3}}{64}\, \eta^4 +\frac{{\mu}^{2}\epsilon}{32}\left( 4\,\eta^2 \eta_{xx}+3 \, \eta \eta_x^2 + 6 D_x^{-1}\left({\eta_x^3} \right)\right)\\
 +\frac{\mu\,{\epsilon}^{2}}{1440}\left(504\,\eta\eta_{xxxx}+ 1091\, \eta_x\eta_{xxx}+ 652\, \eta_{xx}^2\right)
 +{\frac {61\, {\epsilon}^{3} }{1890}} \eta_{xxxxxx} \, . \label{w}
\end{multline}
Here $D_x^{-1}$ denotes integration. Assuming $ {\eta_x^3}\rightarrow 0$ rapidly enough as $x\rightarrow -\infty$ one can set $D_x^{-1}\left({\eta_x^3} \right)=\int_{-\infty}^x \eta_x^3\, dx$. The first line of the expansion (\ref{w}) one can find in \cite{whitham74}, terms $f_{2n}$  were derived in \cite{merSm1990} and \cite{johnson2000}, here we extend the expansion to the terms $f_{3n}$ of order $\mu^n\epsilon^m,\ n+m=3$.

Substitution of (\ref{w}) in either (\ref{bc1}) or (\ref{bc2}) leads to equation 
\begin{multline}
0 =\eta_{{t}}+\eta_{{x}}+\frac{3\mu}{2}\eta_{{x}}\eta+ \frac{\epsilon}{6}\,\eta_{{xxx}}  \\-\frac{3{\mu}^{2}}{8}\,{\eta}^{2}\eta_{{x}} 
 + \epsilon\,\mu\left( {\frac{5}{12}}\,\eta\,\eta_{{xxx}}+{\frac {23}{24}}\,\eta_{{x}}\eta_{{xx}} \right)  
+ {\frac{19{\epsilon}^{2}}{360}}\,\eta_{{xxxxx}}    \\+\frac{3{\mu}^{3}}{16}\,{\eta}^{3}\eta_{{x}}
 + \epsilon\,{\mu}^{2}\left(  {\frac {23}{16}}  \eta\,\eta_{{x}}\eta_{{xx}}+  {\frac {5}{16}} {\eta}^{2}\eta_{{xxx}}+  {\frac {19}{32}} {\eta_{{x}}}^{3}  \right)   \\
 + {\epsilon}^{2}\mu \left( {\frac {1079}{1440}}\,\eta_{{xxxx}}\eta_{{x}}+{\frac {317}{288}}\,\eta_{{xx}}\eta_{{xxx}}+{\frac {19}{80}}\,\eta_{{xxxxx}}\eta \right) 
 +  {\frac {55 {\epsilon}^{3}}{3024}}\,\eta_{{xxxxxxx}}.  \label{kdv}
\end{multline}
The first line of this expansion is the standard Korteweg--de Vries equation, corrections in the second line
is the well known result (see \cite{merSm1990}, \cite{johnsonsolution}). For the purpose of our paper  we proceed to the terms of order  $\mu^n\epsilon^m,\ n+m=3$.
\subsection{Asymptotic near-identity transformation}
Following \cite{dullinetal} we shall apply the Galilean and asymptotically invertible near-identity transformations to equation (\ref{kdv}). The purpose of these transformations is to bring the first two lines of equation (\ref{kdv}) in the form, which can be re-scaled to equation (\ref{cceq0}).

First we apply the Galilean transformation
\begin{equation}\label{gal}
 X = x - \delta t,\qquad  T =  t,
\end{equation}
where $\delta$ is a constant which will be determined later. Then we perform 
the Kodama transformation 
\begin{equation}
\eta(X,T) =u  +\mu \left( \alpha_{{1
}} u^{2}+\alpha_{{2}} u _X D_X^{-1}(u)  \right) 
+\epsilon\beta u_{XX} 
\end{equation}
to a new dependent variable $u=u(X,T)$. And finally we apply the Helmholtz operator ${\cal H} = 1 - \epsilon\gamma \partial_X^2$ to the equation obtained.

In the Galilean transformation and the Helmholtz operator we set $\delta=9/19$ and $\gamma = 19/60$ in order to vanish the coefficients at the terms $u_{XXX}$ and $u_{XXXXX}$ respectively. The choice   
$
  \alpha_1 = 7/20,\  \alpha_2 = -1/5,\
\beta = 1/30
$
excludes the term $u^2 u_X$ and guarantees (see details in \cite{dullinetal}) that the first line of the resulting equation 
\begin{multline}\label{transformed}
0= u_{{T}}+{\frac {10}{19}}\,u_{{X}}+  \frac{3\,\mu}{2}\,\,u  u_{{X}} -{\frac {19\, \epsilon}{60}}\,u_{{TXX}}
-\frac{\mu\epsilon}{120}\left(38\,u_{{XX}}u_{{X}}+19\,u_{{XXX}}u \right) \\
+ {\frac {223\,\epsilon^3}{151200}}\,u_{{XXXXXXX}}-{\frac {3\,\mu^3}{100}}\,u_{{XX}}D^{-1}_X \left( u  \right) \left(  u^{2}-2\,u_{{X}}D^{-1}_X \left( u  \right) \right) \\
+\frac{\mu^2\epsilon}{2400}\left( {976}\,u  u_{{X}}u_{{XX}}-48uu_{{XXXX}} D^{-1}_X \left( u  \right) +48\,{u_{{XX}}}^{2}D^{-1}_X \left( u  \right)+680u_{{XXX}}  u^{2}\right. \\
\left. +2765\,{u_{{X}}}^{3} \right)
+\frac{\mu\epsilon^2}{3600}\left( 903\,u_{{XXXX}}u_{{X}}+316 u_{{XXXXX}}u  +305u_{{XX}}u_{{XXX}}\right)
\end{multline}
can be re-scaled to (\ref{cceq0}).

Until this stage the asymptotic theory is consistent and  equation (\ref{transformed}) is asymptotically equivalent to the KdV expansion. The first line of (\ref{transformed}) was derived in \cite{dullinetal}, our contribution is in the retaining
of the next corrections in the asymptotic expansion (the last three lines in (\ref{transformed})). 

\section{Derivation of the Camassa-Holm equation and its inconsistency with the asymptotic expansion}

Analysing the publications with derivations of the Camassa-Holm equation as a long wave asymptotic expansion we notice that they have a similar pattern. Starting from the Euler equation or a certain well established model of water waves (the Green-Naghdi model, the generalised Serre model, etc) the authors arrive to the equation similar (up to an inessential re-scaling with constant coefficients) to the first line of (\ref{transformed}). Then they truncate the expansion at this level and  re-scale it to the form (\ref{cceq0}). We shall do the same re-scaling, but accounting the next correction.

The most general re-scaling of variables that transforms the first line of equation (\ref{transformed}) into (\ref{cceq0}) is 
\begin{equation}\label{rescaling}
  u=\,A\,U, \qquad y =  \frac{2}{19}\,{\frac {\sqrt {285}}{\sqrt {\epsilon}}} X ,\qquad \tau =\frac{1}{19}\,{\frac {\mu A\sqrt {285}}{\sqrt {\epsilon}}}\,T,
\end{equation}
where $A$ is an arbitrary constant and $\omega= 10/(19\,A\,\mu)$. The re-scaling (\ref{rescaling}) balances the terms in the the equation by eliminating the small parameters $\epsilon$ and $\mu$ (except the the term with $\omega$). It is easy to see that after this re-scaling the small parameters disappear from the correction (the last three lines in (\ref{transformed})), which takes the form  
\begin{multline}
\label{correcswithomega}
{\frac {12}{361\,{{\omega}^{2}}}}\,{U_{{yy}}U_{{y}} \left( D^{-1}_y\left( U  \right) \right) ^{2}}-{\frac {6}{361\,{{\omega}^{2}}}}\,U  ^{2} {U_{{yy}} D^{-1}_y\left( U  \right)}
+{\frac {2440}{361\,{\omega}}}\,{U_{{y}}U  U_{{yy}}}\\
+{\frac {2765}{722\,{\omega}}}\, {(U_{{y}})^{3}}
+{\frac {24}{361\,{\omega}}}\, {(U_{{yy}})^{2}D^{-1}_y\left( U  \right)}
-{\frac {24}{361\,{\omega}}}\, {U U_{{yyyy}}D^{-1}_y\left( U  \right)}
+{\frac {340}{361\,{\omega}}}\,{U^{2}U_{{yyy}}}\\
+{\frac {1806}{361}}\,U_{{y}}U_{{yyyy}}
+{\frac {610}{361}}\,U_{{yy}}U_{{yyy}}+{\frac {632}{361}}\,U  U_{{yyyyy}}+{\frac {446}{2527}}\,\omega\,U_{{yyyyyyy}}\, .
\end{multline}\\
One can also demonstrate that the small parameters disappear from all higher corrections.

One could hope that for the exact soliton or peakon solutions the correction term (\ref{correcswithomega}) and all higher corrections vanish. It is obviously not the case for peakons, moreover the constant $\omega$, which has to be set zero (for the existence of peakons) is in the denominator. It does not happen with solitons either. To show that we used the exact soliton solution of equation (\ref{cceq0}) taken from \cite{johnsonsolution}:
\[
 U(y,\tau) =\frac{\left( c-2\,\omega \right) }{ 1+2 \omega c^{-1}\sinh^2(\theta)} 
\]
where $c$ is an arbitrary constant satisfying condition $c>2\omega$ and $\theta$ is a function of $y-c\tau$ implicitly given by equation 
\[
 y-c\tau =\frac{2 \theta}{ \sqrt{1-2\omega c^{-1}}}+
 \ln  \left( \frac{\cosh  \left(\theta-{\rm arctanh}  \sqrt{1-2\omega c^{-1}} \,\right)  }{ \cosh  \left(\theta+{\rm arctanh}  \sqrt{1-2\omega c^{-1}}\, \right)} \right)\, .
\]

\begin{figure}
\centering
$\begin{array}{ccc}
{\includegraphics[scale=0.2]{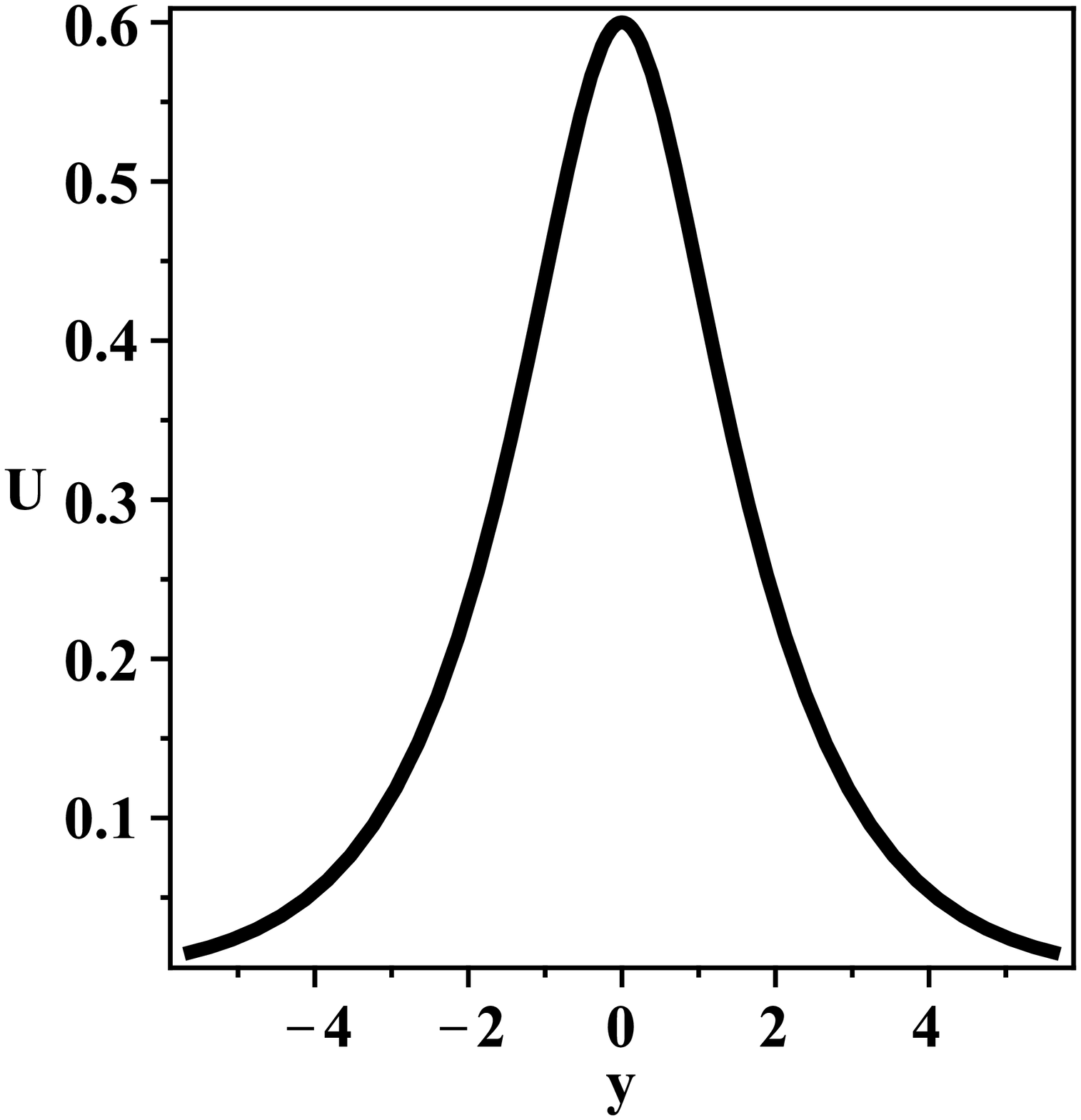}} &
{\includegraphics[scale=0.2]{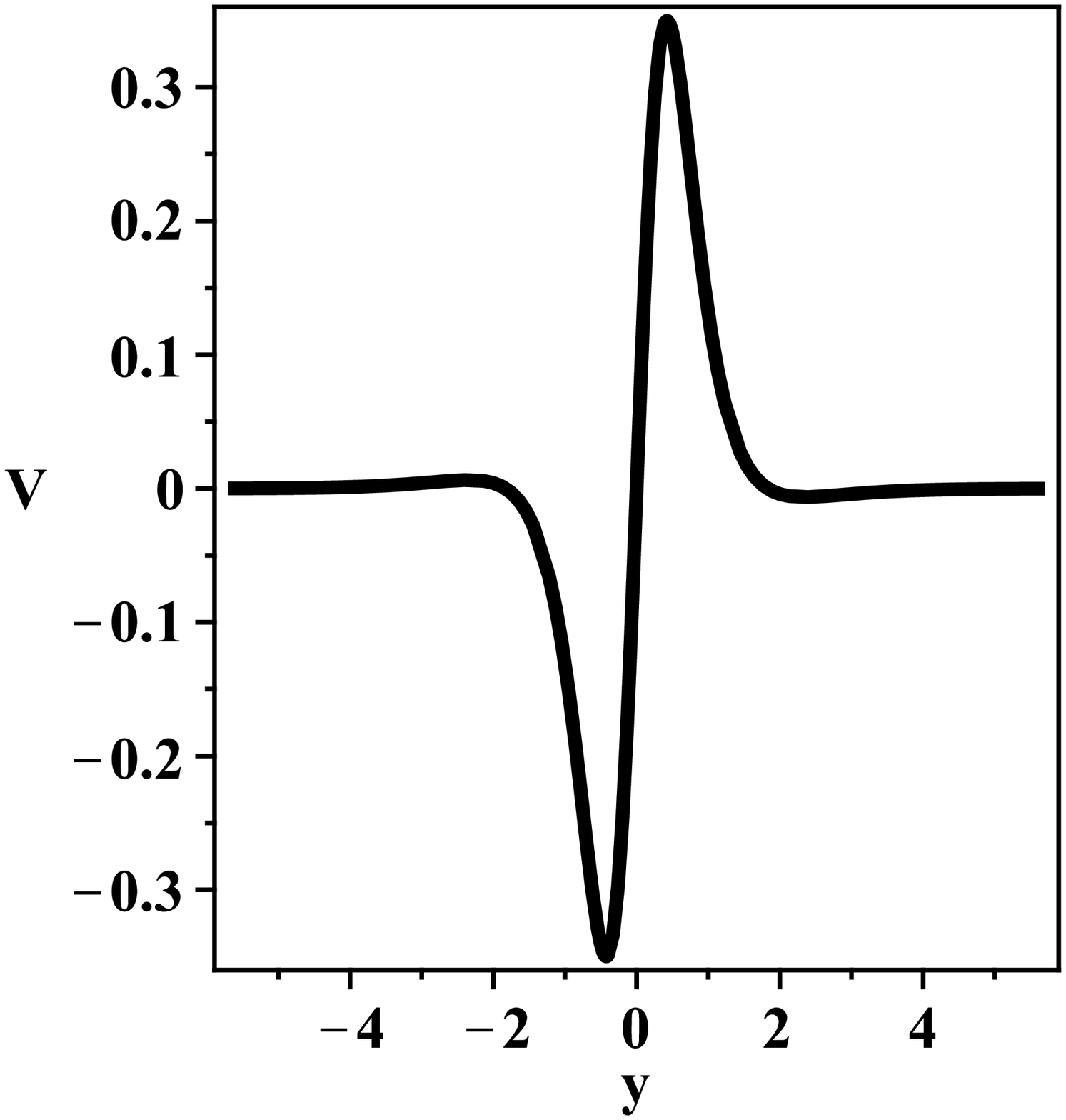}} &
{\includegraphics[scale=0.2]{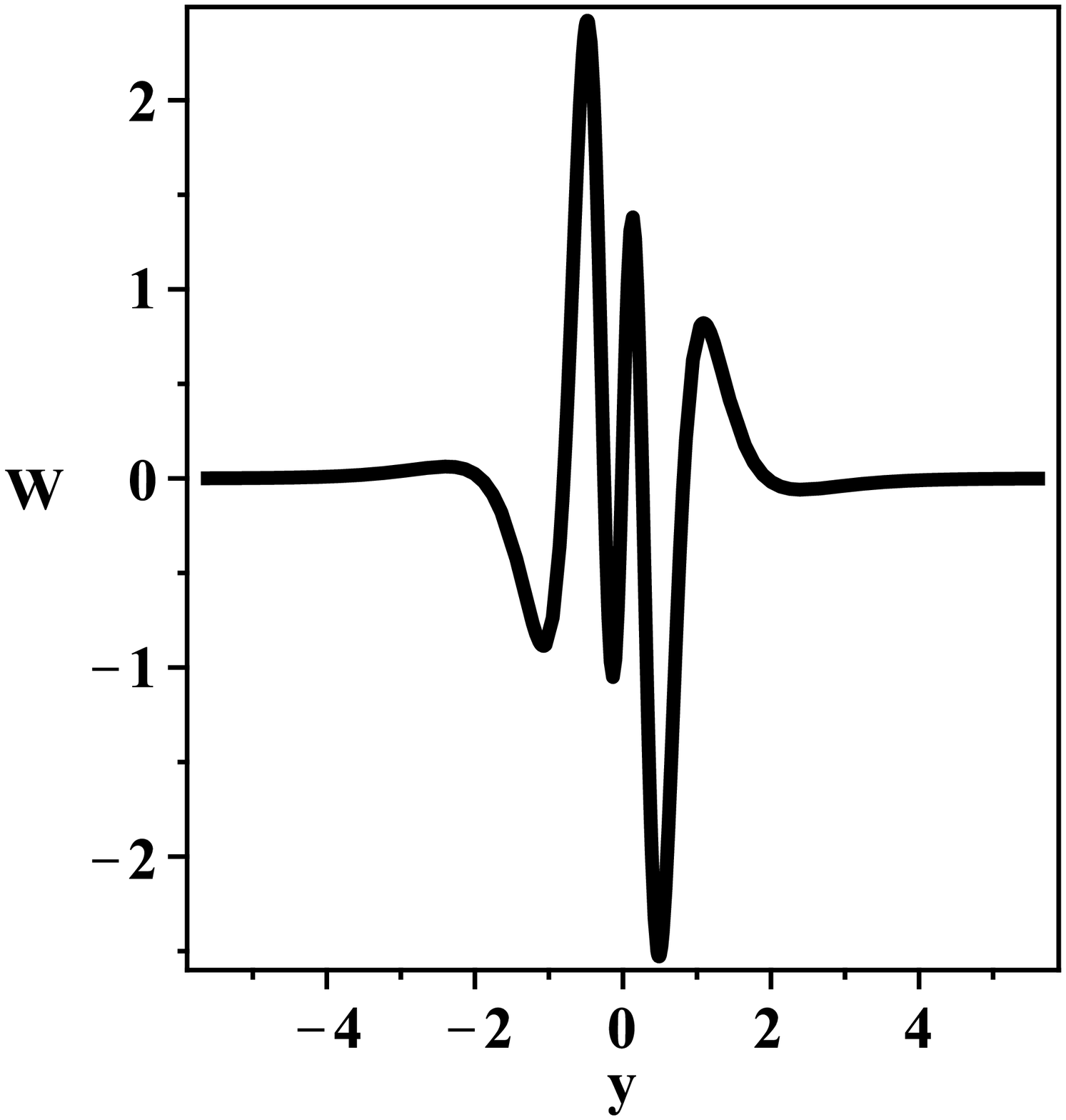}} 
\end{array}$
\caption{(From left to right) The soliton shape ($c=1,\, \omega=0.2$), the value of $UU_{yyy}$ term evaluated on this soliton solution, the value of the correction (\ref{correcswithomega}).}
\end{figure}

Taking $c=1,\ \omega=0.2$ we compared a contribution from one of the  terms of the Camassa-Holm equation with the value of the correction (\ref{correcswithomega}). It is shown on Fig.1 that the correction (\ref{correcswithomega}) is much bigger then a contribution from the term of the equation (\ref{cceq0}) evaluated on the soliton solution with this choice of parameters and this fact does not depend on the choice of the term.  

The re-scaling (\ref{rescaling}) is a basic error, which leads to the inconsistency with the long wave asymptotic theory. Indeed, suppose we have started from  equation (\ref{cceq0}) and have found its solution of a size or characteristic wavelength $\lambda\sim 1$. Re-scaling this solution to the variable $X$ we find from (\ref{rescaling}) that the size in this variable is $\Lambda=\frac{19\sqrt{\epsilon}}{2\sqrt{285}}\lambda\approx 0.56\sqrt{\epsilon}\lambda $. 
Since the Galilean transformation (\ref{gal}) does not change the scale, the wave has the same size in the variable $x$. Coming back to the physical dimensional variable $x'$ we realise that the size of the wave is $\lambda'=L\Lambda\approx 0.56 \lambda h_0 $ which is in contradiction with the long wave assumption $h^2_0/(\lambda')^2\approx 3.2/\lambda^2 \ll 1$.

One can consider solutions of (\ref{cceq0}) of an extremely large characteristic length $\lambda\gg 1$, but with the   same accuracy such solutions can be described by the Korteweg--de Vries equation with the first correction (\ref{kdv}) (whose integration is not much different from the KdV itself ,\cite{kodama1}, \cite{kodama2}). For $\lambda\gg 1$ it follows from \cite{holmdull} that equations    
(\ref{cceq0}) and (\ref{kdv}) are asymptotically equivalent, but the theory of the latter is much simpler and well developed.

\section{Conclusion}

The Camassa-Holm model for shallow water waves does not represent the long-wave asymptotic. Any solution of this model, which is not asymptotically close to a solution of the Korteweg--de Vries equation is an artefact of the model and irrelevant to the theory of shallow water waves. In the literature there are many papers with implicit criticism of various aspects of the derivation of the Camassa-Holm model and its validity as well as attempts to rectify them, but  they only contribute further to the confusion. Serious concerns about the asymptotic sense of the Camassa-Holm equation as a water wave theory has been raised in \cite{johnson2000}.  In our paper we put an end to desperate attempts to justify the Camassa-Holm model using the long wave asymptotic theory.  Based on our consideration it is not difficult to conclude, that neither the Camassa-Holm model nor the Degasperis-Procesi equation (\cite{degpro}) can represent a long wave asymptotic in problems of Hydrodynamic, Physics of Condensed Matter, Plasmas, etc, and that the peakon solutions are irrelevant for the long wave asymptotic theory.   

Having said so, we do not want to undermine the mathematical value of equation (\ref{cceq0}). It is a fine example of an integrable multi-Hamiltonian system with interesting  associated spectral theory, unusual properties, etc. It may have other applications due to the discovery by \cite{misiolek} (see also \cite{misiolekKhesin}) that equation (\ref{cceq0}) is the Euler equation for the geodesic flow on the Virasoro group with respect to the right-invariant Sobolev $H^1$ -metric. As a mathematical object, the class of integrable equations discovered by Fuchssteiner and Fokas is a valuable contribution to the theory of differential equations.

\section*{Acknowledgements} We would like to thank D.~D.~Holm for sending us his papers with a detailed derivation of the Camassa-Holm model. We are grateful to the participants of Leeds Applied Nonlinear Dynamics seminar, in particularly C.~A.~Jones and D.~W.~Hughes, for discussion of this work, useful questions and advises. AVM presented\footnote{See  ${ http://zakharov70.itp.ac.ru/report/day\_2/07\_mikhailov/mikhailov.pdf}$ for the talk slides.} the result of this paper on The Fifth International Workshop
``Solitons, Collapses and Turbulence:
Achievements, Developments and Perspectives'',
Chernogolovka,  Russia and he is grateful to the participants of the Workshop including E.~A.~Kuznetsov, V.~E.~Zakharov, A.~C.~Newell, A.~S.~Fokas and F.~Dias for helpful comments and discussion.

\end{document}